\documentclass[aps,preprintnumbers,eqsecnum,amsmath,amssymb,showpacs,nofootinbib]{revtex4}
\usepackage{graphicx}
\usepackage{dcolumn}
\usepackage{bm}
\usepackage{epsfig}

\begin{document}
\title{Extraction of ground-state decay constant from dispersive sum rules:
\\
QCD vs potential models}
\author{Wolfgang Lucha$^{a}$, Dmitri Melikhov$^{a,b,c}$, and Silvano Simula$^{d}$}
\affiliation{
$^a$HEPHY, Austrian Academy of Sciences, Nikolsdorfergasse 18, A-1050, Vienna, Austria\\
$^b$Faculty of Physics, University of Vienna, Boltzmanngasse 5, A-1090 Vienna, Austria\\
$^c$SINP, Moscow State University, 119991, Moscow, Russia\\
$^d$INFN, Sezione di Roma III, Via della Vasca Navale 84, I-00146, Roma, Italy}
\date{\today}
\begin{abstract}
We compare the extraction of the ground-state decay constant from the two-point correlator in QCD and in potential models and show
that the results obtained at each step of the extraction procedure follow a very similar pattern. 
We prove that allowing for a Borel-parameter-dependent effective continuum threshold yields two essential improvements compared 
to employing a Borel-parameter-independent quantity: 
(i) It reduces considerably the (unphysical) dependence of the extracted bound-state mass and the decay constant on the
Borel parameter. 
(ii) In a potential model, where the actual value of the~decay constant is known from the Schr\"odinger equation, a
Borel-parameter-dependent threshold leads to an improvement of the accuracy of the extraction procedure.
Our findings suggest that in QCD a Borel-parameter dependent threshold leads to a more reliable and accurate determination of  
bound-state characteristics by the method of sum rules. 
\end{abstract}
\maketitle
\section{Introduction}
In a series of recent publications we studied the extraction of the ground-state parameters from SVZ sum rules~\cite{svz} (see
also, e.g., \cite{nsvz,rad,ck}). We made use of a quantum-mechanical potential model since this is essentially the
only case where the standard procedures adopted in the method of sum rules may be tested:  
the estimates for the ground-state parameters obtained by these procedures may be compared with the actual 
values of the ground-state parameters calculated from the Schr\"odinger equation, 
thus providing an unambiguous check of the reliability of the method.

The main results of our papers may be summarized as follows: 
(i) The standard approximation of a constant effective continuum threshold does not allow one to probe the accuracy of the
extracted hadron parameter \cite{lms_2ptsr,lms_3ptsr,m_lcsr,lms_scalar}. 
(ii) Allowing for a Borel-parameter-dependent effective continuum threshold (we denote the Borel parameter $\tau$ in QCD and 
$T$ in the potential model) and fixing this quantity by using the information on the ground-state mass leads to a considerable 
improvement of the accuracy of the method \cite{lmss}.

The goal of this letter is to demonstrate that the results obtained at each step of the extraction procedure 
both in QCD and in potential models follow the same pattern. This similarity gives a strong argument that all our findings
concerning the extraction of bound-state parameters from correlators obtained in potential model apply also to QCD. In
particular, it points out a way of improving the results for bound-state parameters obtained from various correlators in QCD.

The paper is organized as follows:
In Section \ref{sect:qcd} we recall the QCD results from Ref.~\cite{jamin} for the vacuum correlator of two pseudoscalar
currents --- the basic object for the extraction of $f_B$ within the framework of QCD sum rules. Section \ref{sect:qm} provides the
analogous results for a quantum-mechanical model for the case of a potential containing confining and Coulomb interactions. Section
\ref{sect:sr} compares the procedures of extracting the decay constant. Section \ref{sect:conclusions} summarizes our conclusions.

\section{\label{sect:qcd}Correlator and sum rule in QCD}
Let us consider the correlator
\begin{eqnarray}
\label{Pi_QCD} \Pi(p^2)=i \int d^4x e^{ipx}\langle
0|T\left(j_5(x)j_5^\dagger(0)\right)| 0\rangle
\end{eqnarray}
of two pseudoscalar currents
$j_5(x)=(m_b+m_u)\bar q(x) i\gamma_5 b(x)$.
The Borel-transformed operator product expansion (OPE) series for this correlator has the form
\begin{eqnarray}
\label{OPE_QCD}
\Pi(\tau)=\int\limits^\infty_{(m_b+m_u)^2}e^{-s\tau}\rho_{\rm pert}(s,\mu)ds + \Pi_{\rm power}(\tau,\mu),
\end{eqnarray}
where the perturbative spectral density reads 
\begin{eqnarray}
\label{rhopert} \rho_{\rm pert}(s,\mu)=\rho^{(0)}(s,\mu)+\frac{\alpha_s(\mu)}{\pi}\rho^{(1)}(s,\mu)+
\left(\frac{\alpha_s(\mu)}{\pi}\right)^2\rho^{(2)}(s,\mu)+\cdots, 
\end{eqnarray}
$\mu$ being the renormalization scale. 
We make use of the results for $\rho_{\rm pert}$ reported in \cite{jamin} and do not reproduce the explicit expression for this
quantity here. Following the argument of \cite{jamin} we work in terms of the running masses in the $\overline{\rm MS}$ scheme.
Therefore, in all expressions in this section the quark masses $m_b$ and $m_u$, and $\alpha_s$ are the $\overline{\rm MS}$
running quantities at the scale~$\mu$. Recall that the full Borel-transformed correlator (\ref{OPE_QCD}) does not depend on
the renormalization scale $\mu$; however, both the perturbative expansion truncated to a fixed order in $\alpha_s$ and the
truncated power corrections depend on $\mu$. We provide numerical estimates for $\mu=m_b$; for this choice of the scale the known
terms of the perturbative expansion exhibit a good hierarchy. We set $m_b(m_b)=4.2$ GeV and for other QCD parameters make use of
the central values reported in Table I of \cite{jamin}.\footnote{It is well known that the numerical value
of the correlator depends sizeably on the values of $m_b(m_b)$ and on the specific choice of the renormalization scale $\mu$.
However, a discussion of this dependence is far beyond the scope of this paper. A detailed analysis of $f_B$ in QCD is deferred to a 
separate publication.} 
The power corrections have been also considered in \cite{jamin}:
\begin{eqnarray}
\label{power_QCD} &&\Pi_{\rm power}(\tau,\mu=m_b)=
(m_b+m_u)^2e^{-m_b^2\tau} \\ \nonumber
&&\hspace{1cm}\times\left\{-m_b\langle \bar qq\rangle \left[
1+\frac{2C_F\alpha_s}{\pi}\left(1-\frac{m_b^2\tau}2\right)
-(1+m_b^2\tau)\frac{m_u}{2m_b}+\frac{m_u^2}{2} m_b^2\tau^2
+\frac{m_0^2\tau}{2}\left(1-\frac{m_b^2\tau}{2}\right) \right]+
\frac{1}{12}\langle{\frac{\alpha_s}{\pi} FF}\rangle \right\}.
\end{eqnarray}
The parameter $m_0^2$ describes the contribution of the four-quark condensate. Notice that radiative corrections to the condensates
increase rather fast with $\tau$.

The correlator (\ref{Pi_QCD}) may be calculated in terms of hadron intermediate states:
\begin{eqnarray}
\label{hadron} \Pi(\tau)=\Pi_{\rm g}(\tau)+\mbox{contributions of excited states}, 
\qquad \Pi_{\rm g}(\tau)\equiv {f_B^2M_B^4}e^{-M_B^2\tau},
\end{eqnarray}
where $f_B$ is the decay constant of the $B$-meson, defined by
\begin{eqnarray}
\label{decay_constant}
(m_b+m_u)\langle 0 |\bar u i\gamma_5 b| B \rangle = f_B M_B^2.
\end{eqnarray}
For large values of $\tau$ the contributions of the excited states decrease faster than the ground-state contribution and $\Pi(\tau)$
is dominated by the ground state. Unfortunately, the truncated OPE does not allow to evaluate the correlator~at sufficiently large
$\tau$, so the excited states give a sizeable contribution to $\Pi(\tau)$ for the considered values of $\tau$. 

According to the duality assumption, the contribution of the excited states is described by the perturbative 
contribution above some effective continuum threshold $s_{\rm eff}$. Then one obtains the following relation:
\begin{eqnarray}
\label{SR_QCD}
\Pi_{\rm g}(\tau)=\Pi_{\rm dual}(\tau, s_{\rm eff})
\end{eqnarray}
with 
\begin{eqnarray}
\Pi_{\rm dual}(\tau, s_{\rm eff})\equiv
\int\limits^{s_{\rm eff}(\tau)}_{(m_b+m_u)^2} e^{-s\tau}\rho_{\rm pert}(s,\mu)\,ds + \Pi_{\rm power}(\tau,\mu).
\end{eqnarray}
In the region near the physical continuum threshold at $s=(M_{B^*}+m_\pi)^2$, the perturbative spectral density and the
hadron spectral density are very different. Consequently, the effective continuum threshold as defined in (\ref{SR_QCD}) turns
out to be necessarily a function of the Borel parameter $\tau$.

The necessity of the $\tau$-dependence of $s_{\rm eff}$ may be understood by comparing the left-hand side (l.h.s.) and right-hand
side (r.h.s.) of (\ref{SR_QCD}): the only way to obtain a single exponential on the l.h.s.\ for a given spectral density of the
integral representation on the r.h.s.\ is to have a $\tau$-dependent $s_{\rm eff}$.

The $\tau$-dependence of $s_{\rm eff}$ may be also demonstrated explicitly: for any value of the ground-state parameters on the
l.h.s.\ of (\ref{SR_QCD}) one can obtain numerically $s_{\rm eff}$ and see that it does depend on $\tau$.

One should be aware of the fact that the $\tau$-dependence of $s_{\rm eff}$ cannot and does not contradict to any principles of
field theory: the dual correlator is a {\it hand-made} object; such an object does not emerge in field theory. Therefore, the
properties of the dual correlator (e.g., its analytic properties) are very different from the properties of the field-theoretic
correlators.

Clearly, the standard assumption of a $\tau$-independent $s_{\rm eff}$ is a possible assumption. We shall demonstrate, however,
that relaxing this assumption leads to a visible improvement of the obtained results.

We define the dual decay constant and the dual invariant mass by the relations
\begin{eqnarray}
\label{fdual}
f_{\rm dual}^2(\tau)=M_B^{-4} e^{M_B^2\tau}\Pi_{\rm dual}(\tau, s_{\rm eff}(\tau)), \qquad
\label{mdual}
M_{\rm dual}^2(\tau)=-\frac{d}{d\tau}\log \Pi_{\rm dual}(\tau, s_{\rm eff}(\tau)).
\end{eqnarray}
Notice that the deviation of the dual mass from the actual mass of the ground state gives an indication of the excited-state
contributions picked up by the dual correlator.

\section{\label{sect:qm}Correlator and sum rule in potential models}
In parallel to QCD, let us consider a quantum-mechanical model with a potential containing a confining part, for which we take
the HO form, and an attractive Coulomb interaction:
\begin{eqnarray}
\label{H}
H=\frac{k^2}{2m}+\frac{m\omega^2 r^2}{2}-\frac{\alpha}{r}.
\end{eqnarray}
A quantum-mechanical analogue of the Borelized two-point function has the form \cite{nsvz}
\begin{eqnarray}
\Pi(T)=\langle \vec r'=0|\exp(-H T)|\vec r=0\rangle.
\end{eqnarray}
We construct the analogue of the OPE series for this correlator by retaining, similar to the QCD case, the perturbative contributions
up to $O(\alpha^2)$ (three loops of the non-relativistic field theory) and two power corrections, including $O(\alpha)$
corrections to them. The resulting expression reads (the correlator for a pure Coulomb potential can be found
in~\cite{voloshin}):\footnote{Interestingly, at small values of $T$ the system behaves like a free system, since the contribution
of the confining potential as well as~that of the radiative corrections vanish for small $T$. According to \cite{svz} this is
a signature of asymptotic freedom. So, a non-relativistic potential model (\ref{H}) even with a constant $\alpha$ behaves
like an asymptotically free theory.}
\begin{eqnarray}
\Pi_{\rm OPE}(T)&=&\Pi_{\rm pert}(T)+\Pi_{\rm power}(T),
\nonumber\\
\Pi_{\rm pert}(T)&=&
\left(\frac{m}{2\pi T}\right)^{3/2}
\left[1+\sqrt{2\pi mT}\alpha+\frac{1}{3}m\pi^2 T \alpha^2\right],
\nonumber\\
\Pi_{\rm power}(T)&=&
\left(\frac{m}{2\pi T}\right)^{3/2}
\left[-\frac{1}{4}\omega^2 T^2\left(1+\frac{11}{12}\sqrt{2\pi m T}\alpha\right)
+\frac{19}{480}\omega^4 T^4\left(1+\frac{1541}{1824}\sqrt{2\pi m T}\alpha\right)\right].
\end{eqnarray}
Now, according to the standard procedures of the method of sum rules, the dual correlator is obtained as follows: we represent
the perturbative contribution as a single spectral representation in the relative kinetic energy $z$ of the interacting quarks and
cut this representation at $z_{\rm eff}$:
\begin{eqnarray}
\Pi_{\rm dual}(T,z_{\rm eff})=
\left(\frac{m}{2\pi}\right)^{3/2}\int\limits_0^{z_{\rm eff}} dz
\exp(-z T)\left[2\sqrt{\frac{z}{\pi}}+\sqrt{2\pi m}\alpha+\frac{\pi^{3/2} m\alpha^2}{3\sqrt{z}}\right] +\Pi_{\rm power}(T).
\end{eqnarray}
By construction, the dual correlator is related to the ground-state contribution by
\begin{eqnarray}
\label{Pidual}
\Pi_{\rm dual}(T,z_{\rm eff})=\Pi_{\rm g}(T)\equiv R_{\rm g}\exp(-E_{\rm g}T),\qquad R_{\rm g}=|\psi_{\rm g}(r=0)|^2.
\end{eqnarray}
As we have shown in our previous studies of potential models, the effective continuum threshold defined according to (\ref{Pidual})
is a function of the Borel time parameter $T$.

For our numerical analysis, we adopt the following parameter values: a reduced quark mass of $m=0.175$~GeV,~which corresponds
to a constituent quark mass of 0.350~GeV relevant for nonrelativistic computations; $\omega=0.5$~GeV, which leads to a
realistic radius of the $q\bar q$ system, and $\alpha=0.3$.

The energy and wave function of the ground state are found by solving numerically the Schr\"odinger equation with the help of
the Mathematica code provided in \cite{franz}: the ground-state energy is $E_g=0.6473$~GeV [for comparison, a pure HO model yields
$E^{\rm HO}_g=0.75$~GeV]; the ground-state wave function at the origin has the value $\psi(r=0)=0.0783$~GeV$^{3/2}$ 
[$\psi^{\rm HO}(r=0)=(m \omega/\pi)^{3/4}= 0.068$ GeV$^{3/2}$]. 
Obviously, the effect of the Coulomb interaction is not small.

\section{\label{sect:sr}Extraction of the decay constant}

Whether a $\tau$-independent or some $\tau$-dependent effective continuum threshold is considered, the crucial problem~is~the
choice of the criterion for fixing this quantity. We shall proceed as follows:
\subsection{The Borel window}
First, we must fix the working $\tau$ window where, on the one hand, the OPE gives an accurate description of the exact
correlator (i.e., the higher-order radiative and power corrections are small) and, on the other hand, the ground~state gives a
sizeable contribution to the correlator. In QCD, we set the window as follows:
\begin{eqnarray}
\label{our_window}
0.05\, {\rm GeV}^{-2} < \tau < 0.18\, {\rm GeV}^{-2}.
\end{eqnarray}
In the region $\tau < 0.18$ GeV$^{-2}$ the $\alpha_s$ and $\alpha_s^2$ terms to $\Pi(\tau)$ contribute less than 10\% and
3\% of the leading term, respectively. Power corrections give about 20\% of the leading term. We point out that the radiative
corrections to the condensates increase rather fast with $\tau$, so it is preferable to stay at relatively low values of $\tau$.
Therefore our window is located at the lower values of $\tau$ compared to the window adopted in \cite{jamin}; in the region 
(\ref{our_window}) the accuracy of the truncated OPE is higher. 

It is known that the experimental value of the $B$ meson decay constant is $f_B\approx 200$ MeV. Adopting this value, we may
calculate the relative contribution of the ground state to the correlator. In our window it does not exceed 50\%.

In the potential model, we choose the window as 
\begin{eqnarray}
0.2\, {\rm GeV}^{-1}< T < 0.8\, {\rm GeV}^{-1}.
\end{eqnarray}
For these values of $T$ the omitted unknown higher-order power corrections are negligible, so the correlator is known with good
accuracy. The relative contribution of the ground state to the correlator amounts to 10\% at $T=0.2\, {\rm GeV}^{-1}$ and to 50\%
at $T=0.8\,{\rm GeV}^{-1}$.

We shall see that for a relative ground-state contribution of this size our procedure allows one to extract the decay constant with a
reasonable accuracy. 
\subsection{Fixing the effective continuum threshold}
Widely used is the so-called stability criterion: one looks for that constant value of $s_{\rm eff}$ for which the extracted~decay
constant is most stable in the window. This Borel stability is an implementation of a self-evident statement that the physical
observable cannot depend on $\tau$, an auxiliary parameter of the method. The problem is, however, that the independence of a hadron
decay constant of $\tau$, being a necessary condition, is not sufficient to guarantee the extraction of the right value. We have
given several examples for potential models \cite{lms_2ptsr,lms_3ptsr} which nicely demonstrate that
assuming~a $\tau$-independent effective continuum threshold and fixing its value by requiring maximal stability in the Borel
window may lead to the extraction of a very inaccurate value.

In this paper, we consider a different algorithm for the extraction of $f_B,$ which makes use of the knowledge of the
ground-state mass \cite{jamin}. In parallel to QCD we present also the results for a quantum-mechanical model (\ref{H}). 
This is done in order to demonstrate the way our algorithm works in the case where the
exact value of the decay constant is known. A comparison makes~clear that, with respect to the
extraction procedure, there are no essential differences between QCD and quantum mechanics.

The algorithm developed in our previous works and established to work well for different correlators in the potential model is very
simple: we consider a set of $\tau$-dependent Ans\"atze for the effective continuum threshold (for the case of the potential model
one just replaces $\tau\to T$ and $M\to E$):
\begin{eqnarray}
\label{zeff} s^{(n)}_{\rm eff}(\tau)=
\sum\limits_{j=0}^{n}s_j^{(n)}\tau^{j}.
\end{eqnarray}
Obviously, the standard $\tau$-independent effective continuum threshold is also taken account by (\ref{zeff}). Now, we fix the
parameters on the r.h.s.\ of (\ref{zeff}) as follows: we calculate the dual mass squared according to (\ref{mdual}) for the
$\tau$-dependent $s_{\rm eff}$ of Eq.~(\ref{zeff}). We then evaluate $M^2_{\rm dual}(\tau)$ at several values of $\tau =
\tau_i$ ($i = 1,\dots, N$, where $N$ can be taken arbitrarily large) chosen uniformly in the window. Finally, we minimize the
squared difference between $M^2_{\rm dual}$ and the known value $M^2_B$:
\begin{eqnarray}
\label{chisq}
\chi^2 \equiv \frac{1}{N} \sum_{i = 1}^{N} \left[ M^2_{\rm dual}(\tau_i) - M_B^2 \right]^2.
\end{eqnarray}
This gives us the parameters of the effective continuum thresholds. As soon as the latter are fixed, it is straightforward
to obtain the decay constant.

\begin{figure}
\begin{tabular}{cc}
\includegraphics[width=6cm]{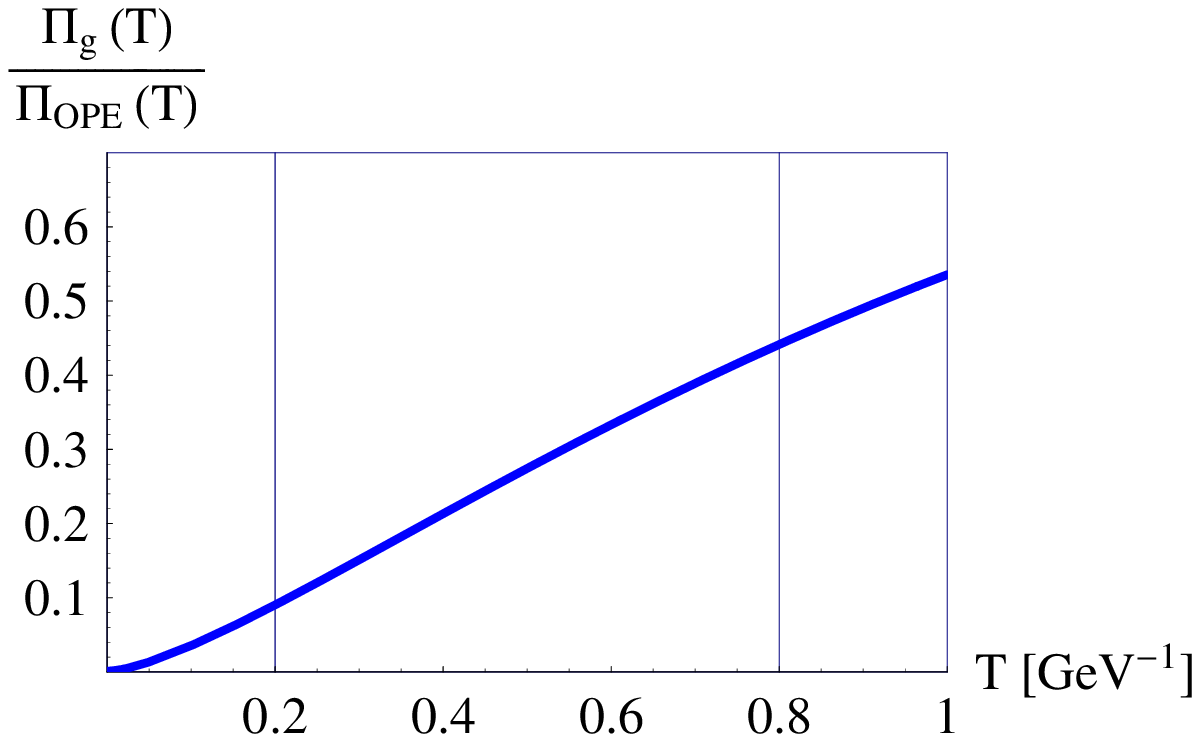} & \includegraphics[width=6cm]{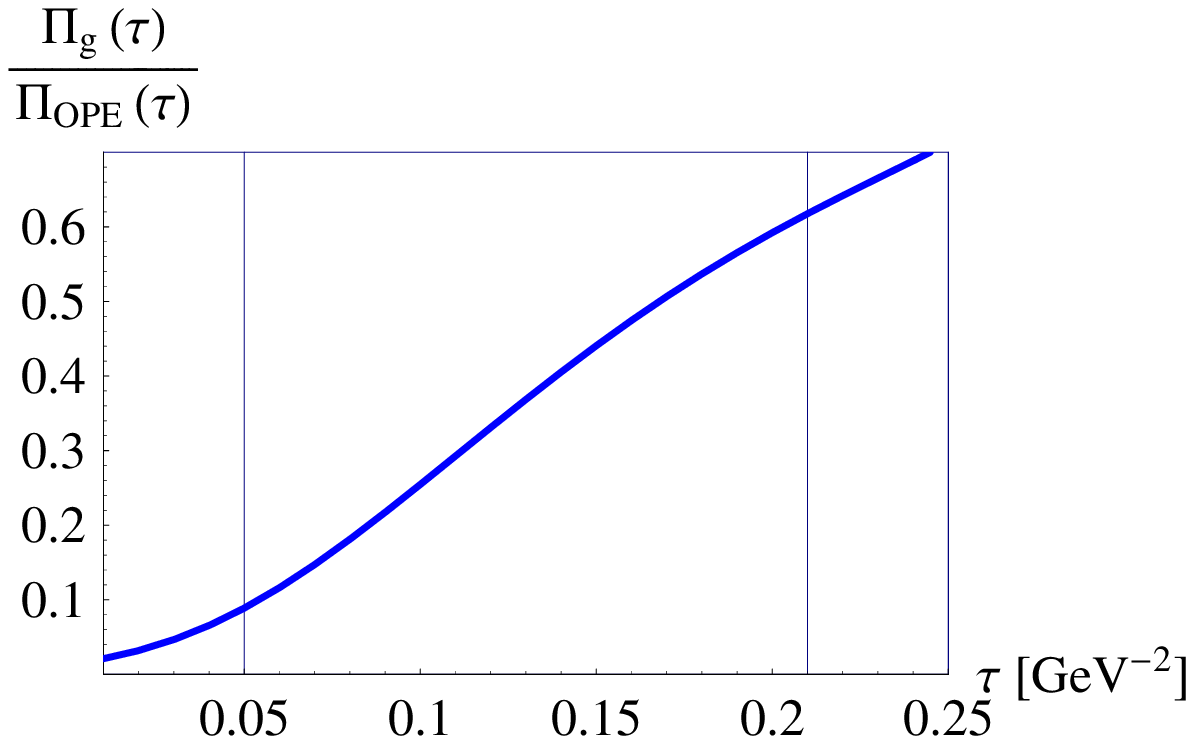}\\
(a) & (b) \\
\includegraphics[width=7cm]{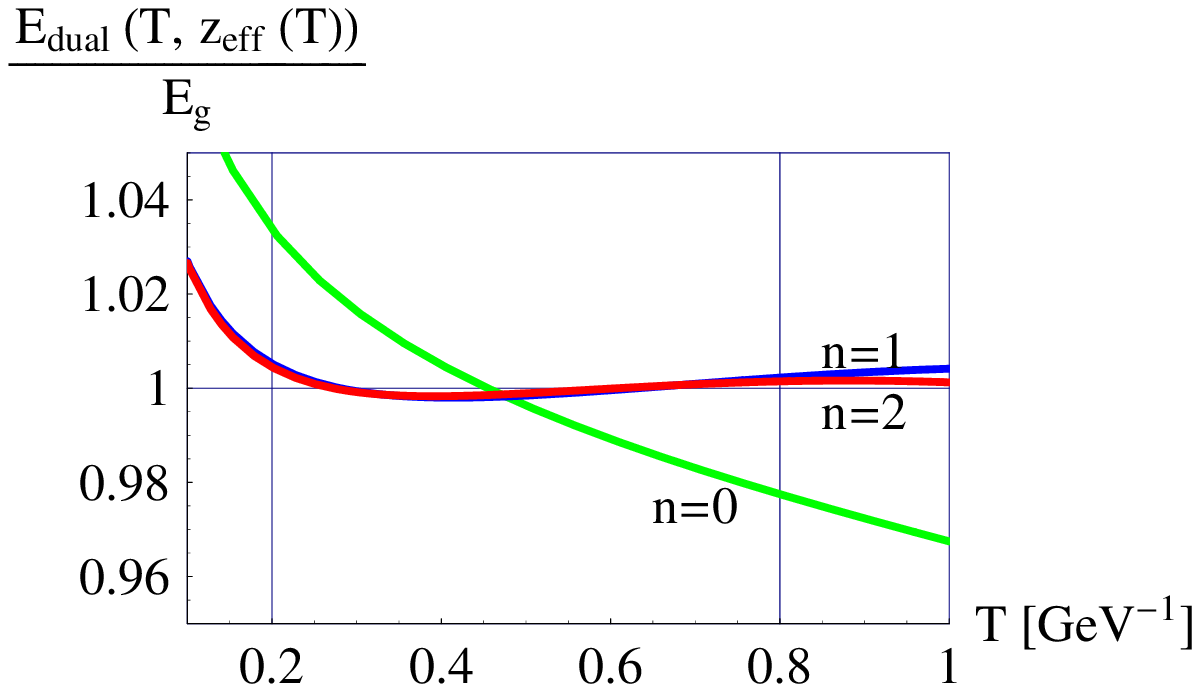} & \includegraphics[width=7cm]{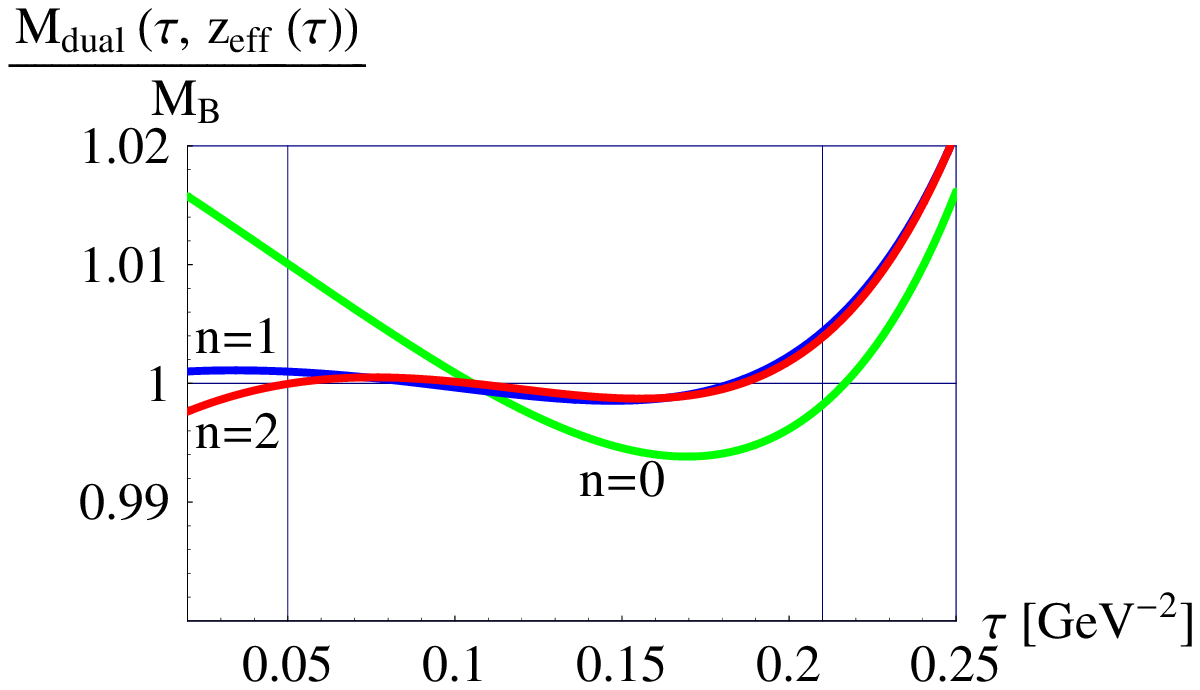}\\
(c) & (d) \\
\includegraphics[width=7cm]{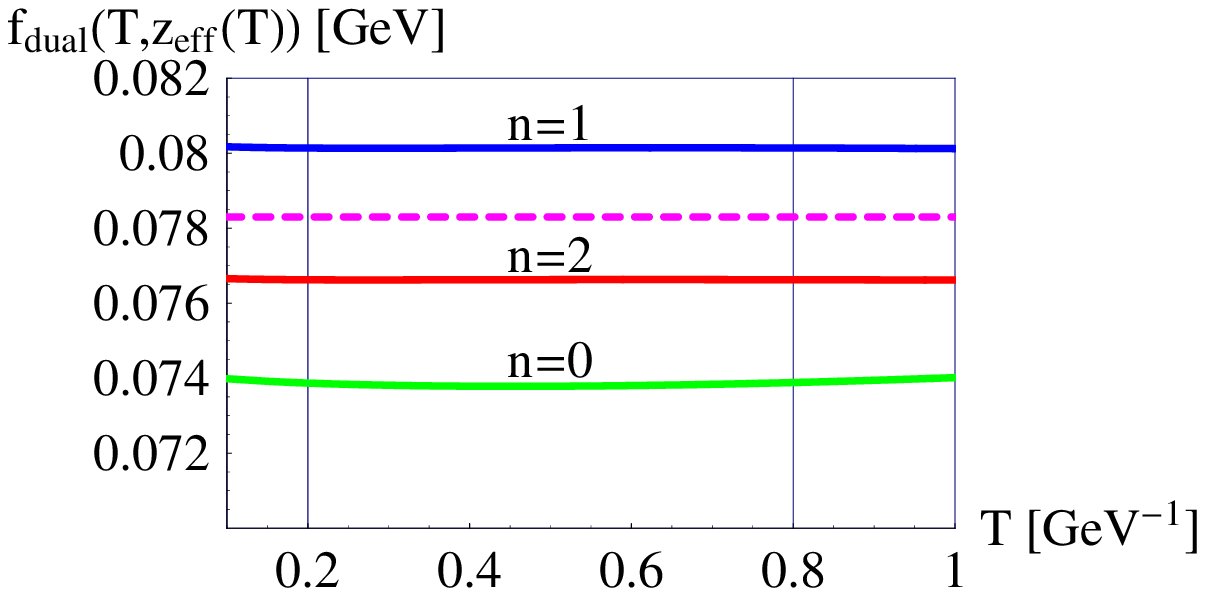} & \includegraphics[width=7cm]{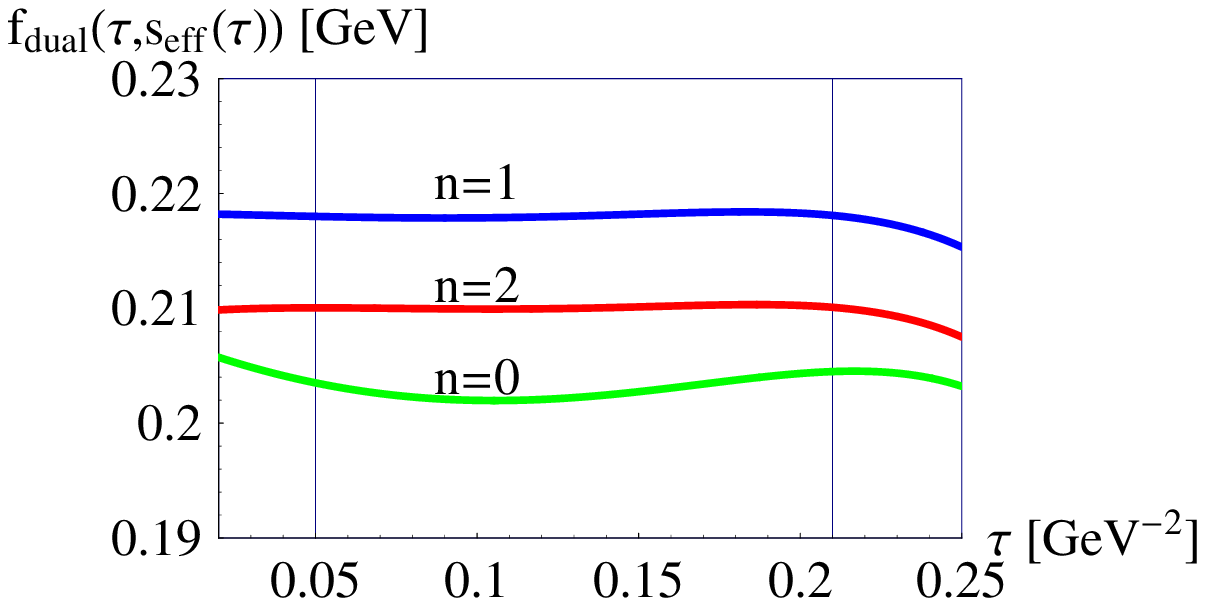}\\
(e) & (f)
\end{tabular}
\caption{\label{Plot:1} Left column: potential model (\ref{H}); right column: QCD. 
First line: relative contribution of the ground state to the
correlator; second line: fitted dual mass; third line: corresponding dual decay constant. The dashed line in Fig.~(e)
corresponds to the~true value of the decay constant obtained by solving the Schr\"odinger equation. The index $n$ is the power of
the polynomial Ansatz~for the Borel-parameter-dependent effective continuum threshold.}
\end{figure}

Figure~\ref{Plot:1} shows the results for QCD and for our potential model; in the latter the actual value of the decay
constant has been found from the Schr\"odinger equation, so that we may control each step of the extraction procedure.

First, let us notice that the Borel-parameter-dependent effective thresholds corresponding to $n=1$ and $n=2$ lead to a visible 
improvement of the stability of the dual mass [Figs.~\ref{Plot:1}(c) and (d)] compared to the constant threshold. 
This means that the dual correlator for $n>0$ is less contaminated by the excited states; according to the 
philosophy of QCD sum rules the  
better stability of $M_{\rm dual}$ with $n>0$ is an important achievement for the trustability of the results.

According to Fig.~\ref{Plot:1}, in the potential model the true value of the decay constant lies in the band provided by the
linear ($n=1$) and the quadratic ($n=2$) fits. We have checked that this result holds in a broad range of the parameters of the
potential model. The similarities of each step of the extraction procedure in QCD and in the potential model~are
evident.\footnote{One may observe that outside the window the behavior of the dual mass in QCD and in the potential model is not
exactly the same. This is related to the fact that the quark condensate in QCD is negative, whereas the corresponding power
correction in potential models (for any confining potential) has a positive sign.} Therefore, it is tempting to expect that also in
QCD the decay constant lies in the range provided by the linear and the quadratic fits. Anyway, the difference of the results
obtained for $n=1$ and $n=2$ constitutes a realistic estimate of the intrinsic uncertainty of the extracted decay constant. 
If one considers only the standard constant Ansatz ($n=0$) for the effective continuum threshold, the accuracy of the
extracted decay constant cannot be probed.

\section{\label{sect:conclusions}Discussion and conclusions}
We have presented a detailed analysis of the extraction of the decay constant from the two-point function in QCD and in a
potential model. Our results may be summarized as follows:
\begin{itemize}
\item[(i)]
The comparison presented in this work makes obvious that, with respect to the extraction of the ground-state parameters, there
are no essential differences between QCD and quantum mechanics: as soon as the parameters of the Lagrangian are fixed, and the
truncated OPE is calculated with a reasonable accuracy (taking into account also the relevant choice of the renormalization scale
in QCD), the extraction procedures are very similar.

At first glance, this similarity might look surprising since we know that the structure of bound states in potential models and in
QCD are rather different. However, the method of dispersive sum rules does not make use of (and does not provide) information about
the details of the ground-state structure. What really matters for extracting the ground-state parameters in this method is the
structure of the OPE for a given correlator. Since the structure of the OPE in QCD and of its analogue in potential
models is rather similar, it should not be surprising at all that the extraction procedures in potential models and in
QCD are similar, too.

In view of the above similarity, our previous results for the extraction of the ground-state parameters (including also the form
factors \cite{lmss}) obtained in potential models have direct implications for the corresponding analyses~in QCD and should be
taken quite seriously.

\item[(ii)]
Allowing for $\tau$-dependent Ans\"atze for the effective continuum threshold leads to two essential improvements:

(a) The stability of both the dual mass and the dual decay constant in the window is considerably improved if one proceeds
from the standard $\tau$-independent to the $\tau$-dependent Ansatz for the effective continuum threshold.

(b) In the potential model, where the exact decay constant has been calculated from the Schr\"odinger equation, allowing for a
$\tau$-dependent effective continuum threshold and fixing its parameters according to 
(\ref{chisq}) ---
i.e., by minimizing the deviation of the dual mass from the known ground-state mass in the window --- leads to the extraction of a
more accurate value. As follows from our analysis, a realistic band of values of the decay constant is provided by the numerical
results obtained with the linear and quadratic Ans\"atze for the effective continuum threshold. The intrinsic uncertainty (i.e.,
the one related to the extraction procedure) of the decay constant found in this way is expected to be at the level of a few percent.

Although not rigorous in the mathematical sense, this estimate for the systematic uncertainty may be considered as a realistic 
educated guess supported by findings in models where the true value of the decay constant is known. Moreover, we have~doubts that a 
more rigorous estimate of the 
intrinsic error of the method of sum rules may be obtained in principle.
\end{itemize}

\vspace{.5cm}
\noindent {\it Acknowledgements}: We are grateful to Matthias Jamin for providing us with his Mathematica code for the
calculation of $f_B$. D.~M.\ gratefully acknowledges financial support from the Austrian Science Fund (FWF) under project P20573
and from the Alexander von Humboldt-Stiftung. D.~M. expresses his gratitude to the Institute of Theoretical Physics of the
Heidelberg University for hospitality during his visit, where this work was started. The work was supported in part by Federal Agency
for Science and Innovation of Russian Federation under state contract 02.740.11.0244 and by EU Contract No.
MRTN-CT-2006-035482, ``FLAVIAnet''.

\end{document}